
  \input miniltx
  \def\Gin@driver{pdftex.def}
  \input color.sty
  \input graphicx.sty
  \resetatcatcode

%
%

%
%
%
%

\def\Serif{cmr}
\def\SerifBold{cmbx}
\def\SerifItalics{cmti}
\def\SerifSlanted{cmsl}
\def\SerifBoldItalics{cmbxti}
\def\SansSerif{cmss}
\def\SansSerifBold{cmssbx}
\def\SansSerifItalics{cmssi}
\def\SansSerifSlanted{cmssi}
\def\Math{cmmi}
\def\Symbols{cmsy}
\def\MathBold{cmmib}
\def\MoreSymbols{cmex}
\def\Typewriter{cmtt}
\def\Gothic{eufm}
\def\Double{msbm}
\def\Relazioni{msam}

= 			\Serif10 			at 5pt
= 		\SerifBold10 		at 5pt
= 	\SerifItalics10 	at 5pt
=		\SerifSlanted10 	at 5pt
=	\SerifBoldItalics10	at 5pt
= 		\SansSerif10 		at 5pt
=	\SansSerifBold10	at 5pt
=	\SansSerifItalics10	at 5pt
=	\SansSerifSlanted10	at 5pt
=				\Math10				at 5pt
=			\MathBold10			at 5pt
=			\Symbols10			at 5pt
=		\MoreSymbols10		at 5pt
=		\Typewriter10		at 5pt
=			\Gothic10			at 5pt
=			\Double10			at 5pt

= 			\Serif10 			at 7pt
= 		\SerifBold10 		at 7pt
= 	\SerifItalics10 	at 7pt
=	\SerifSlanted10 	at 7pt
=\SerifBoldItalics10	at 7pt
= 		\SansSerif10 		at 7pt
= 	\SansSerifBold10 	at 7pt
=\SansSerifItalics10	at 7pt
=\SansSerifSlanted10	at 7pt
=			\Math10				at 7pt
=		\MathBold10			at 7pt
=			\Symbols10			at 7pt
=		\MoreSymbols10		at 7pt
=		\Typewriter10		at 7pt
=			\Gothic10			at 7pt
=			\Double10			at 7pt

= 			\Serif10 			at 8pt
= 		\SerifBold10 		at 8pt
= 	\SerifItalics10 	at 8pt
=	\SerifSlanted10 	at 8pt
=\SerifBoldItalics10	at 8pt
= 		\SansSerif10 		at 8pt
= 	\SansSerifBold10 	at 8pt
=\SansSerifItalics10 at 8pt
=\SansSerifSlanted10 at 8pt
=			\Math10				at 8pt
=		\MathBold10			at 8pt
=			\Symbols10			at 8pt
=		\MoreSymbols10		at 8pt
=		\Typewriter10		at 8pt
=			\Gothic10			at 8pt
=			\Double10			at 8pt

= 			\Serif10 			at 10pt
= 		\SerifBold10 		at 10pt
= 		\SerifItalics10 	at 10pt
=		\SerifSlanted10 	at 10pt
=	\SerifBoldItalics10	at 10pt
= 		\SansSerif10 		at 10pt
= 	\SansSerifBold10 	at 10pt
= 	\SansSerifItalics10 at 10pt
= 	\SansSerifSlanted10 at 10pt
=				\Math10				at 10pt
=			\MathBold10			at 10pt
=			\Symbols10			at 10pt
=		\MoreSymbols10		at 10pt
=		\Typewriter10		at 10pt
=			\Gothic10			at 10pt
=			\Double10			at 10pt
=			\Relazioni10			at 10pt

= 				\Serif10 			at 12pt
= 			\SerifBold10 		at 12pt
= 		\SerifItalics10 	at 12pt
=		\SerifSlanted10 	at 12pt
=	\SerifBoldItalics10	at 12pt
= 			\SansSerif10 		at 12pt
= 		\SansSerifBold10 	at 12pt
= 	\SansSerifItalics10 at 12pt
= 	\SansSerifSlanted10 at 12pt
=				\Math10				at 12pt
=			\MathBold10			at 12pt
=			\Symbols10			at 12pt
=		\MoreSymbols10		at 12pt
=			\Typewriter10		at 12pt
=				\Gothic10			at 12pt
=				\Double10			at 12pt

= 			\Serif10 			at 14pt
= 		\SerifBold10 		at 14pt
= 	\SerifItalics10 	at 14pt
=		\SerifSlanted10 	at 14pt
=	\SerifBoldItalics10	at 14pt
= 		\SansSerif10 		at 14pt
= 	\SansSerifBold10 	at 14pt
= \SansSerifSlanted10 at 14pt
= \SansSerifItalics10 at 14pt
=				\Math10				at 14pt
=			\MathBold10			at 14pt
=			\Symbols10			at 14pt
=		\MoreSymbols10		at 14pt
=		\Typewriter10		at 14pt
=			\Gothic10			at 14pt
=			\Double10			at 14pt

\def\NormalStyle{\parindent=5pt\parskip=3pt\normalbaselineskip=14pt%
\def\nt{\tenSerif}%
\def\rm{\fam0\tenSerif}%
\textfont0=\tenSerif\scriptfont0=\sevenSerif\scriptscriptfont0=\fiveSerif
\textfont1=\tenMath\scriptfont1=\sevenMath\scriptscriptfont1=\fiveMath
\textfont2=\tenSymbols\scriptfont2=\sevenSymbols\scriptscriptfont2=\fiveSymbols
\textfont3=\tenMoreSymbols\scriptfont3=\sevenMoreSymbols\scriptscriptfont3=\fiveMoreSymbols
\textfont\itfam=\tenSerifItalics\def\it{\fam\itfam\tenSerifItalics}%
\textfont\slfam=\tenSerifSlanted\def\sl{\fam\slfam\tenSerifSlanted}%
\textfont\ttfam=\tenTypewriter\def\tt{\fam\ttfam\tenTypewriter}%
\textfont\bffam=\tenSerifBold%
\def\bf{\fam\bffam\tenSerifBold}\scriptfont\bffam=\sevenSerifBold\scriptscriptfont\bffam=\fiveSerifBold%
\def\cal{\tenSymbols}%
\def\greekbold{\tenMathBold}%
\def\gothic{\tenGothic}%
\def\Bbb{\tenDouble}%
\def\LieFont{\tenSerifItalics}%
\nt\normalbaselines\baselineskip=14pt%
}

\def\TitleStyle{\parindent=0pt\parskip=0pt\normalbaselineskip=15pt%
\def\nt{\fourteenSansSerifBold}%
\def\rm{\fam0\fourteenSansSerifBold}%
\textfont0=\fourteenSansSerifBold\scriptfont0=\tenSansSerifBold\scriptscriptfont0=\eightSansSerifBold
\textfont1=\fourteenMath\scriptfont1=\tenMath\scriptscriptfont1=\eightMath
\textfont2=\fourteenSymbols\scriptfont2=\tenSymbols\scriptscriptfont2=\eightSymbols
\textfont3=\fourteenMoreSymbols\scriptfont3=\tenMoreSymbols\scriptscriptfont3=\eightMoreSymbols
\textfont\itfam=\fourteenSansSerifItalics\def\it{\fam\itfam\fourteenSansSerifItalics}%
\textfont\slfam=\fourteenSansSerifSlanted\def\sl{\fam\slfam\fourteenSerifSansSlanted}%
\textfont\ttfam=\fourteenTypewriter\def\tt{\fam\ttfam\fourteenTypewriter}%
\textfont\bffam=\fourteenSansSerif%
\def\bf{\fam\bffam\fourteenSansSerif}\scriptfont\bffam=\tenSansSerif\scriptscriptfont\bffam=\eightSansSerif%
\def\cal{\fourteenSymbols}%
\def\greekbold{\fourteenMathBold}%
\def\gothic{\fourteenGothic}%
\def\Bbb{\fourteenDouble}%
\def\LieFont{\fourteenSerifItalics}%
\nt\normalbaselines\baselineskip=15pt%
}

\def\PartStyle{\parindent=0pt\parskip=0pt\normalbaselineskip=15pt%
\def\nt{\fourteenSansSerifBold}%
\def\rm{\fam0\fourteenSansSerifBold}%
\textfont0=\fourteenSansSerifBold\scriptfont0=\tenSansSerifBold\scriptscriptfont0=\eightSansSerifBold
\textfont1=\fourteenMath\scriptfont1=\tenMath\scriptscriptfont1=\eightMath
\textfont2=\fourteenSymbols\scriptfont2=\tenSymbols\scriptscriptfont2=\eightSymbols
\textfont3=\fourteenMoreSymbols\scriptfont3=\tenMoreSymbols\scriptscriptfont3=\eightMoreSymbols
\textfont\itfam=\fourteenSansSerifItalics\def\it{\fam\itfam\fourteenSansSerifItalics}%
\textfont\slfam=\fourteenSansSerifSlanted\def\sl{\fam\slfam\fourteenSerifSansSlanted}%
\textfont\ttfam=\fourteenTypewriter\def\tt{\fam\ttfam\fourteenTypewriter}%
\textfont\bffam=\fourteenSansSerif%
\def\bf{\fam\bffam\fourteenSansSerif}\scriptfont\bffam=\tenSansSerif\scriptscriptfont\bffam=\eightSansSerif%
\def\cal{\fourteenSymbols}%
\def\greekbold{\fourteenMathBold}%
\def\gothic{\fourteenGothic}%
\def\Bbb{\fourteenDouble}%
\def\LieFont{\fourteenSerifItalics}%
\nt\normalbaselines\baselineskip=15pt%
}

\def\ChapterStyle{\parindent=0pt\parskip=0pt\normalbaselineskip=15pt%
\def\nt{\fourteenSansSerifBold}%
\def\rm{\fam0\fourteenSansSerifBold}%
\textfont0=\fourteenSansSerifBold\scriptfont0=\tenSansSerifBold\scriptscriptfont0=\eightSansSerifBold
\textfont1=\fourteenMath\scriptfont1=\tenMath\scriptscriptfont1=\eightMath
\textfont2=\fourteenSymbols\scriptfont2=\tenSymbols\scriptscriptfont2=\eightSymbols
\textfont3=\fourteenMoreSymbols\scriptfont3=\tenMoreSymbols\scriptscriptfont3=\eightMoreSymbols
\textfont\itfam=\fourteenSansSerifItalics\def\it{\fam\itfam\fourteenSansSerifItalics}%
\textfont\slfam=\fourteenSansSerifSlanted\def\sl{\fam\slfam\fourteenSerifSansSlanted}%
\textfont\ttfam=\fourteenTypewriter\def\tt{\fam\ttfam\fourteenTypewriter}%
\textfont\bffam=\fourteenSansSerif%
\def\bf{\fam\bffam\fourteenSansSerif}\scriptfont\bffam=\tenSansSerif\scriptscriptfont\bffam=\eightSansSerif%
\def\cal{\fourteenSymbols}%
\def\greekbold{\fourteenMathBold}%
\def\gothic{\fourteenGothic}%
\def\Bbb{\fourteenDouble}%
\def\LieFont{\fourteenSerifItalics}%
\nt\normalbaselines\baselineskip=15pt%
}

\def\SectionStyle{\parindent=0pt\parskip=0pt\normalbaselineskip=13pt%
\def\nt{\twelveSansSerifBold}%
\def\rm{\fam0\twelveSansSerifBold}%
\textfont0=\twelveSansSerifBold\scriptfont0=\eightSansSerifBold\scriptscriptfont0=\eightSansSerifBold
\textfont1=\twelveMath\scriptfont1=\eightMath\scriptscriptfont1=\eightMath
\textfont2=\twelveSymbols\scriptfont2=\eightSymbols\scriptscriptfont2=\eightSymbols
\textfont3=\twelveMoreSymbols\scriptfont3=\eightMoreSymbols\scriptscriptfont3=\eightMoreSymbols
\textfont\itfam=\twelveSansSerifItalics\def\it{\fam\itfam\twelveSansSerifItalics}%
\textfont\slfam=\twelveSansSerifSlanted\def\sl{\fam\slfam\twelveSerifSansSlanted}%
\textfont\ttfam=\twelveTypewriter\def\tt{\fam\ttfam\twelveTypewriter}%
\textfont\bffam=\twelveSansSerif%
\def\bf{\fam\bffam\twelveSansSerif}\scriptfont\bffam=\eightSansSerif\scriptscriptfont\bffam=\eightSansSerif%
\def\cal{\twelveSymbols}%
\def\bg{\twelveMathBold}%
\def\gothic{\twelveGothic}%
\def\Bbb{\twelveDouble}%
\def\LieFont{\twelveSerifItalics}%
\nt\normalbaselines\baselineskip=13pt%
}

\def\SubSectionStyle{\parindent=0pt\parskip=0pt\normalbaselineskip=13pt%
\def\nt{\twelveSansSerifItalics}%
\def\rm{\fam0\twelveSansSerifItalics}%
\textfont0=\twelveSansSerifItalics\scriptfont0=\eightSansSerifItalics\scriptscriptfont0=\eightSansSerifItalics%
\textfont1=\twelveMath\scriptfont1=\eightMath\scriptscriptfont1=\eightMath%
\textfont2=\twelveSymbols\scriptfont2=\eightSymbols\scriptscriptfont2=\eightSymbols%
\textfont3=\twelveMoreSymbols\scriptfont3=\eightMoreSymbols\scriptscriptfont3=\eightMoreSymbols%
\textfont\itfam=\twelveSansSerif\def\it{\fam\itfam\twelveSansSerif}%
\textfont\slfam=\twelveSansSerifSlanted\def\sl{\fam\slfam\twelveSerifSansSlanted}%
\textfont\ttfam=\twelveTypewriter\def\tt{\fam\ttfam\twelveTypewriter}%
\textfont\bffam=\twelveSansSerifBold%
\def\bf{\fam\bffam\twelveSansSerifBold}\scriptfont\bffam=\eightSansSerifBold\scriptscriptfont\bffam=\eightSansSerifBold%
\def\cal{\twelveSymbols}%
\def\greekbold{\twelveMathBold}%
\def\gothic{\twelveGothic}%
\def\Bbb{\twelveDouble}%
\def\LieFont{\twelveSerifItalics}%
\nt\normalbaselines\baselineskip=13pt%
}

\def\AuthorStyle{\parindent=0pt\parskip=0pt\normalbaselineskip=14pt%
\def\nt{\tenSerif}%
\def\rm{\fam0\tenSerif}%
\textfont0=\tenSerif\scriptfont0=\sevenSerif\scriptscriptfont0=\fiveSerif
\textfont1=\tenMath\scriptfont1=\sevenMath\scriptscriptfont1=\fiveMath
\textfont2=\tenSymbols\scriptfont2=\sevenSymbols\scriptscriptfont2=\fiveSymbols
\textfont3=\tenMoreSymbols\scriptfont3=\sevenMoreSymbols\scriptscriptfont3=\fiveMoreSymbols
\textfont\itfam=\tenSerifItalics\def\it{\fam\itfam\tenSerifItalics}%
\textfont\slfam=\tenSerifSlanted\def\sl{\fam\slfam\tenSerifSlanted}%
\textfont\ttfam=\tenTypewriter\def\tt{\fam\ttfam\tenTypewriter}%
\textfont\bffam=\tenSerifBold%
\def\bf{\fam\bffam\tenSerifBold}\scriptfont\bffam=\sevenSerifBold\scriptscriptfont\bffam=\fiveSerifBold%
\def\cal{\tenSymbols}%
\def\greekbold{\tenMathBold}%
\def\gothic{\tenGothic}%
\def\Bbb{\tenDouble}%
\def\LieFont{\tenSerifItalics}%
\nt\normalbaselines\baselineskip=14pt%
}

\def\AddressStyle{\parindent=0pt\parskip=0pt\normalbaselineskip=14pt%
\def\nt{\eightSerif}%
\def\rm{\fam0\eightSerif}%
\textfont0=\eightSerif\scriptfont0=\sevenSerif\scriptscriptfont0=\fiveSerif
\textfont1=\eightMath\scriptfont1=\sevenMath\scriptscriptfont1=\fiveMath
\textfont2=\eightSymbols\scriptfont2=\sevenSymbols\scriptscriptfont2=\fiveSymbols
\textfont3=\eightMoreSymbols\scriptfont3=\sevenMoreSymbols\scriptscriptfont3=\fiveMoreSymbols
\textfont\itfam=\eightSerifItalics\def\it{\fam\itfam\eightSerifItalics}%
\textfont\slfam=\eightSerifSlanted\def\sl{\fam\slfam\eightSerifSlanted}%
\textfont\ttfam=\eightTypewriter\def\tt{\fam\ttfam\eightTypewriter}%
\textfont\bffam=\eightSerifBold%
\def\bf{\fam\bffam\eightSerifBold}\scriptfont\bffam=\sevenSerifBold\scriptscriptfont\bffam=\fiveSerifBold%
\def\cal{\eightSymbols}%
\def\greekbold{\eightMathBold}%
\def\gothic{\eightGothic}%
\def\Bbb{\eightDouble}%
\def\LieFont{\eightSerifItalics}%
\nt\normalbaselines\baselineskip=14pt%
}

\def\AbstractStyle{\parindent=0pt\parskip=0pt\normalbaselineskip=12pt%
\def\nt{\eightSerif}%
\def\rm{\fam0\eightSerif}%
\textfont0=\eightSerif\scriptfont0=\sevenSerif\scriptscriptfont0=\fiveSerif
\textfont1=\eightMath\scriptfont1=\sevenMath\scriptscriptfont1=\fiveMath
\textfont2=\eightSymbols\scriptfont2=\sevenSymbols\scriptscriptfont2=\fiveSymbols
\textfont3=\eightMoreSymbols\scriptfont3=\sevenMoreSymbols\scriptscriptfont3=\fiveMoreSymbols
\textfont\itfam=\eightSerifItalics\def\it{\fam\itfam\eightSerifItalics}%
\textfont\slfam=\eightSerifSlanted\def\sl{\fam\slfam\eightSerifSlanted}%
\textfont\ttfam=\eightTypewriter\def\tt{\fam\ttfam\eightTypewriter}%
\textfont\bffam=\eightSerifBold%
\def\bf{\fam\bffam\eightSerifBold}\scriptfont\bffam=\sevenSerifBold\scriptscriptfont\bffam=\fiveSerifBold%
\def\cal{\eightSymbols}%
\def\greekbold{\eightMathBold}%
\def\gothic{\eightGothic}%
\def\Bbb{\eightDouble}%
\def\LieFont{\eightSerifItalics}%
\nt\normalbaselines\baselineskip=12pt%
}

\def\RefsStyle{\parindent=0pt\parskip=0pt%
\def\nt{\eightSerif}%
\def\rm{\fam0\eightSerif}%
\textfont0=\eightSerif\scriptfont0=\sevenSerif\scriptscriptfont0=\fiveSerif
\textfont1=\eightMath\scriptfont1=\sevenMath\scriptscriptfont1=\fiveMath
\textfont2=\eightSymbols\scriptfont2=\sevenSymbols\scriptscriptfont2=\fiveSymbols
\textfont3=\eightMoreSymbols\scriptfont3=\sevenMoreSymbols\scriptscriptfont3=\fiveMoreSymbols
\textfont\itfam=\eightSerifItalics\def\it{\fam\itfam\eightSerifItalics}%
\textfont\slfam=\eightSerifSlanted\def\sl{\fam\slfam\eightSerifSlanted}%
\textfont\ttfam=\eightTypewriter\def\tt{\fam\ttfam\eightTypewriter}%
\textfont\bffam=\eightSerifBold%
\def\bf{\fam\bffam\eightSerifBold}\scriptfont\bffam=\sevenSerifBold\scriptscriptfont\bffam=\fiveSerifBold%
\def\cal{\eightSymbols}%
\def\greekbold{\eightMathBold}%
\def\gothic{\eightGothic}%
\def\Bbb{\eightDouble}%
\def\LieFont{\eightSerifItalics}%
\nt\normalbaselines\baselineskip=10pt%
}



%
%


\def\ModeYes{yes}
\def\ModeNo{no}

\def\ModeUndef{undefined}


\def\nx{\noexpand}
\def\ni{\noindent}
\def\newpage{\vfill\eject}

\def\ss{\vskip 5pt}
\def\ms{\vskip 10pt}
\def\bs{\vskip 20pt}

 \def\,{\mskip\thinmuskip}
 \def\!{\mskip-\thinmuskip}
 \def\>{\mskip\medmuskip}
 \def\;{\mskip\thickmuskip}

%
% ... 
%

\def\refsModePost{post}
\def\refsModeAuto{auto}

\def\dbRefsSatusModeOk{ok}
\def\dbRefsSatusModeError{error}
\def\dbRefsSatusModeWarning{warning}


\newcount\BNUM
\BNUM=0

\def\refs{}

\def\SetModePost{\xdef\refsMode{\refsModePost}}			
\SetModePost

\def\dbRefsStatusOk{%
	\xdef\dbRefsStatus{\dbRefsSatusModeOk}%
	\xdef\dbRefsError{\ModeNo}%
	\xdef\dbRefsWarning{\ModeNo}%
	\xdef\dbRefsInfo{\ModeNo}%
}

\def\dbRefs{%
}

\def\dbRefsGet#1{%
	\xdef\found{N}\xdef\ikey{#1}\dbRefsStatusOk%
	\xdef\key{\ModeUndef}\xdef\tag{\ModeUndef}\xdef\tail{\ModeUndef}%
	\dbRefs%
}

\def\NextRefsTag{%
	\global\advance\BNUM by 1%
}
\def\ShowTag#1{{\bf [#1]}}

\def\dbRefsInsert#1#2{%
\dbRefsGet{#1}%
\if\found Y %
   \xdef\dbRefsStatus{\dbRefsSatusModeWarning}%
   \xdef\dbRefsWarning{record is already there}%
   \xdef\dbRefsInfo{record not inserted}%
\else%
   \toks2=\expandafter{\dbRefs}%
   \ifx\refsMode\refsModeAuto \NextRefsTag
    \xdef\dbRefs{%
   	\the\toks2 \nx\xdef\nx\dbx{#1}%
	\nx\ifx\nx\ikey %
		\nx\dbx\nx\xdef\nx\found{Y}%
		\nx\xdef\nx\key{#1}%
		\nx\xdef\nx\tag{\the\BNUM}%
		\nx\xdef\nx\tail{#2}%
	\nx\fi}%
	\global\xdef\refs{\refs \ss\ni[\the\BNUM]\ #2\par}
   \fi%
   \ifx\refsMode\refsModePost 
    \xdef\dbRefs{%
   	\the\toks2 \nx\xdef\nx\dbx{#1}%
	\nx\ifx\nx\ikey %
		\nx\dbx\nx\xdef\nx\found{Y}%
		\nx\xdef\nx\key{#1}%
		\nx\xdef\nx\tag{\ModeUndef}%
		\nx\xdef\nx\tail{#2}%
	\nx\fi}%
   \fi%
\fi%
}

\def\dbRefsEdit#1#2#3{\dbRefsGet{#1}%
\if\found N 
   \xdef\dbRefsStatus{\dbRefsSatusModeError}%
   \xdef\dbRefsError{record is not there}%
   \xdef\dbRefsInfo{record not edited}%
\else%
   \toks2=\expandafter{\dbRefs}%
   \xdef\dbRefs{\the\toks2%
   \nx\xdef\nx\dbx{#1}%
   \nx\ifx\nx\ikey\nx\dbx %
	\nx\xdef\nx\found{Y}%
	\nx\xdef\nx\key{#1}%
	\nx\xdef\nx\tag{#2}%
	\nx\xdef\nx\tail{#3}%
   \nx\fi}%
\fi%
}

\def\bib#1#2{\RefsStyle\dbRefsInsert{#1}{#2}%
	\ifx\dbRefsStatus\dbRefsSatusModeWarning %
		\message{^^J}%
		\message{WARNING: Reference [#1] is doubled.^^J}%
	\fi%
}

\def\ref#1{\dbRefsGet{#1}%
\ifx\found N %
  \message{^^J}%
  \message{ERROR: Reference [#1] unknown.^^J}%
  \ShowTag{??}%
\else%
	\ifx\tag\ModeUndef \NextRefsTag%
		\dbRefsEdit{#1}{\the\BNUM}{\tail}%
		\dbRefsGet{#1}%
		\global\xdef\refs{\refs \ss\ni [\tag]\ \tail\par}
	\fi
	\ShowTag{\tag}%
\fi%
}

\def\ShowBiblio{\ms\Ensure{\SectionEnsure}%
{\SectionStyle\ni References}%
{\RefsStyle\refs}%
}

\newcount\CHANGES
\CHANGES=0
\def\AuxFile{7}
\def\PreventDoubleOn{\xdef\PreventDoubleLabel{\ModeYes}}

\PreventDoubleOn

\def\StoreLabel#1#2{\xdef\itag{#2}
 \ifx\PreModeStatus\ModeNo %
   \message{^^J}%
   \errmessage{You can't use Check without starting with OpenPreMode (and finishing with ClosePreMode)^^J}%
 \else%
   \immediate\write\AuxFile{\nx\dbLabelPreInsert{#1}{\itag}}%
   \dbLabelGet{#1}%
   \ifx\itag\tag %
   \else%
	\global\advance\CHANGES by 1%
 	\xdef\itag{(?.??)}%
    \fi%
   \fi%
}

\def\PreModeStatus{\ModeNo}

\def\ClosePreMode{\immediate\closeout\AuxFile%
  \ifnum\CHANGES=0%
	\message{^^J}%
	\message{**********************************^^J}%
	\message{**  NO CHANGES TO THE AuxFile  **^^J}%
	\message{**********************************^^J}%
 \else%
	\message{^^J}%
	\message{**************************************************^^J}%
	\message{**  PLAEASE TYPESET IT AGAIN (\the\CHANGES)  **^^J}%
    \errmessage{**************************************************^^ J}%
  \fi%
  \edef\PreModeStatus{\ModeNo}%
}

\def\dbLabelSatusModeOk{ok}

\def\dbLabelSatusModeWarning{warning}

\def\dbLabelStatusOk{%
	\xdef\dbLabelStatus{\dbLabelSatusModeOk}%
	\xdef\dbLabelError{\ModeNo}%
	\xdef\dbLabelWarning{\ModeNo}%
	\xdef\dbLabelInfo{\ModeNo}%
}

\def\dbLabel{%
}

\def\dbLabelGet#1{%
	\xdef\found{N}\xdef\ikey{#1}\dbLabelStatusOk%
	\xdef\key{\ModeUndef}\xdef\tag{\ModeUndef}\xdef\pre{\ModeUndef}%
	\dbLabel%
}

\def\ShowLabel#1{%
 \dbLabelGet{#1}%
 \ifx\tag \ModeUndef %
 	\global\advance\CHANGES by 1%
 	(?.??)%
 \else%
 	\tag%
 \fi%
}

\def\dbLabelPreInsert#1#2{\dbLabelGet{#1}%
\if\found Y %
  \xdef\dbLabelStatus{\dbLabelSatusModeWarning}%
   \xdef\dbLabelWarning{Label is already there}%
   \xdef\dbLabelInfo{Label not inserted}%
   \message{^^J}%
   \errmessage{Double pre definition of label [#1]^^J}%
\else%
   \toks2=\expandafter{\dbLabel}%
    \xdef\dbLabel{%
   	\the\toks2 \nx\xdef\nx\dbx{#1}%
	\nx\ifx\nx\ikey %
		\nx\dbx\nx\xdef\nx\found{Y}%
		\nx\xdef\nx\key{#1}%
		\nx\xdef\nx\tag{#2}%
		\nx\xdef\nx\pre{\ModeYes}%
	\nx\fi}%
\fi%
}

\def\dbLabelInsert#1#2{\dbLabelGet{#1}%
\xdef\itag{#2}%
\dbLabelGet{#1}%
\if\found Y %
	\ifx\tag\itag %
	\else%
	   \ifx\PreventDoubleLabel\ModeYes %
		\message{^^J}%
		\errmessage{Double definition of label [#1]^^J}%
	   \else%
		\message{^^J}%
		\message{Double definition of label [#1]^^J}%
	   \fi%
	\fi%
   \xdef\dbLabelStatus{\dbLabelSatusModeWarning}%
   \xdef\dbLabelWarning{Label is already there}%
   \xdef\dbLabelInfo{Label not inserted}%
\else%
   \toks2=\expandafter{\dbLabel}%
    \xdef\dbLabel{%
   	\the\toks2 \nx\xdef\nx\dbx{#1}%
	\nx\ifx\nx\ikey %
		\nx\dbx\nx\xdef\nx\found{Y}%
		\nx\xdef\nx\key{#1}%
		\nx\xdef\nx\tag{#2}%
		\nx\xdef\nx\pre{\ModeNo}%
	\nx\fi}%
\fi%
}


\newcount\PART
\newcount\CHAPTER
\newcount\SECTION
\newcount\SUBSECTION
\newcount\FNUMBER

\PART=0
\CHAPTER=0
\SECTION=0
\SUBSECTION=0	
\FNUMBER=0

\def\LastPart{\ModeUndef}
\def\LastChapter{\ModeUndef}
\def\LastSection{\ModeUndef}
\def\LastSubSection{\ModeUndef}
\def\LastClaim{\ModeUndef}
\def\Last{\ModeUndef}

\newdimen\TOBOTTOM
\newdimen\LIMIT

\def\Ensure#1{\ \par\ \immediate\LIMIT=#1\immediate\TOBOTTOM=\the\pagegoal\advance\TOBOTTOM by -\pagetotal%
\ifdim\TOBOTTOM<\LIMIT\newpage \else%
\vskip-\parskip\vskip-\parskip\vskip-\baselineskip\fi}

\def\PartLabel{\the\PART}
\def\NewPart#1{\global\advance\PART by 1%
         \bs\ni{\PartStyle  Part \PartLabel:}
         \bs\ni{\PartStyle #1}\newpage%
         \CHAPTER=0\SECTION=0\SUBSECTION=0\FNUMBER=0%
         \gdef\Left{#1}%
         \global\edef\Last{\PartLabel}%
         \global\edef\LastPart{\PartLabel}%
         \global\edef\LastChapter{\ModeUndef}%
         \global\edef\LastSection{\ModeUndef}%
         \global\edef\LastSubSection{\ModeUndef}%
         \global\edef\LastClaim{\ModeUndef}}
\def\ChapterLabel{\the\CHAPTER}
\def\NewChapter#1{\global\advance\CHAPTER by 1%
         \bs\ni{\ChapterStyle  Chapter \ChapterLabel: #1}\ms%
         \SECTION=0\SUBSECTION=0\FNUMBER=0%
         \gdef\Left{#1}%
         \global\edef\Last{\ChapterLabel}%
         \global\edef\LastChapter{\ChapterLabel}%
         \global\edef\LastSection{\ModeUndef}%
         \global\edef\LastSubSection{\ModeUndef}%
         \global\edef\LastClaim{\ModeUndef}}
\def\SectionEnsure{3cm}
\def\NewSection#1{\Ensure{\SectionEnsure}\gdef\SectionLabel{\the\SECTION}\global\advance\SECTION by 1%
         \ms\ni{\SectionStyle  \SectionLabel.\ #1}\ss%
         \SUBSECTION=0\FNUMBER=0%
         \gdef\Left{#1}%
         \global\edef\Last{\SectionLabel}%
         \global\edef\LastSection{\SectionLabel}%
         \global\edef\LastSubSection{\ModeUndef}%
         \global\edef\LastClaim{\ModeUndef}}
\def\NewAppendix#1#2{\Ensure{\SectionEnsure}\gdef\SectionLabel{#1}\global\advance\SECTION by 1%
         \bs\ni{\SectionStyle  Appendix \SectionLabel.\ #2}\ss%
         \SUBSECTION=0\FNUMBER=0%
         \gdef\Left{#2}%
         \global\edef\Last{\SectionLabel}%
         \global\edef\LastSection{\SectionLabel}%
         \global\edef\LastSubSection{\ModeUndef}%
         \global\edef\LastClaim{\ModeUndef}}
\def\Acknowledgements{\Ensure{\SectionEnsure}\gdef\SectionLabel{}%
         \ms\ni{\SectionStyle  Acknowledgments}\ss%
         \SECTION=0\SUBSECTION=0\FNUMBER=0%
         \gdef\Left{}%
         \global\edef\Last{\ModeUndef}%
         \global\edef\LastSection{\ModeUndef}%
         \global\edef\LastSubSection{\ModeUndef}%
         \global\edef\LastClaim{\ModeUndef}}
\def\SubSectionEnsure{2cm}
\def\SubSectionLabel{\ifnum\SECTION>0 \the\SECTION.\fi\the\SUBSECTION}
\def\NewSubSection#1{\Ensure{\SubSectionEnsure}\global\advance\SUBSECTION by 1%
         \ms\ni{\SubSectionStyle #1}\ss%
         \global\edef\Last{\SubSectionLabel}%
         \global\edef\LastSubSection{\SubSectionLabel}}
\def\SetNumberingModeN{\def\ClaimLabel{(\the\FNUMBER)}}
\def\SetNumberingModeSN{\def\ClaimLabel{(\ifnum\SECTION>0 \SectionLabel.\fi%
      \the\FNUMBER)}}
\def\SetNumberingModeCSN{\def\ClaimLabel{(\ifnum\CHAPTER>0 \the\CHAPTER.\fi%
      \ifnum\SECTION>0 \SectionLabel.\fi%
      \the\FNUMBER)}}

\def\NewClaim{\global\advance\FNUMBER by 1%
    \ClaimLabel%
    \global\edef\LastClaim{\ClaimLabel}%
    \global\edef\Last{\ClaimLabel}}

\def\HideLabels{\xdef\ShowLabelsMode{\ModeNo}}
\HideLabels

\def\fn{\eqno{\NewClaim}} 
\def\fl#1{%
\ifx\ShowLabelsMode\ModeYes%
 \eqno{{\buildrel{\hbox{\AbstractStyle[#1]}}\over{\hfill\NewClaim}}}%
\else%
 \eqno{\NewClaim}%
\fi%
\dbLabelInsert{#1}{\ClaimLabel}}
\def\fprel#1{\global\advance\FNUMBER by 1\StoreLabel{#1}{\ClaimLabel}%
\ifx\ShowLabelsMode\ModeYes%
\eqno{{\buildrel{\hbox{\AbstractStyle[#1]}}\over{\hfill.\itag}}}%
\else%
 \eqno{\itag}%
\fi%
}

\def\cl#1{\global\advance\FNUMBER by 1\dbLabelInsert{#1}{\ClaimLabel}%
\ifx\ShowLabelsMode\ModeYes%
${\buildrel{\hbox{\AbstractStyle[#1]}}\over{\hfill\ClaimLabel}}$%
\else%
  $\ClaimLabel$%
\fi%
}
\def\cprel#1{\global\advance\FNUMBER by 1\StoreLabel{#1}{\ClaimLabel}%
\ifx\ShowLabelsMode\ModeYes%
${\buildrel{\hbox{\AbstractStyle[#1]}}\over{\hfill.\itag}}$%
\else%
  $\itag$%
\fi%
}

\def\Note{\ms\leftskip 3cm\rightskip 1.5cm\AbstractStyle}
\def\endNote{\par\leftskip 2cm\rightskip 0cm\NormalStyle\ss}


\parindent=7pt
\leftskip=2cm
\newcount\SideIndent
\newcount\SideIndentTemp
\SideIndent=0
\newdimen\SectionIndent
\SectionIndent=-8pt

\def\sidebar{\vrule height15pt width.2pt }
\def\endcorner{\hbox{\hbox{\vrule height6pt width.2pt}\vbox to6pt{\vfill\hbox
to4pt{\leaders\hrule height0.2pt\hfill}}}}
\def\begincorner{\hbox{\hbox{\vrule height6pt width.2pt}\vbox to6pt{\hbox
to4pt{\leaders\hrule height0.2pt\hfill}}}}
\def\endbegincorner{\hbox{\vbox to15pt{\endcorner\vskip-6pt\begincorner\vfill}}}
\def\SideShow{\SideIndentTemp=\SideIndent \ifnum \SideIndentTemp>0 
\loop\sidebar\hskip 2pt \advance\SideIndentTemp by-1\ifnum \SideIndentTemp>1 \repeat\fi}

\def\BeginSection{{\vbadness 100000 \par\ni\hskip\SectionIndent%
\SideShow\vbox to 15pt{\vfill\begincorner}}\global\advance\SideIndent by1\vskip-10pt}

\def\EndSection{{\vbadness 100000 \par\ni\global\advance\SideIndent by-1%
\hskip\SectionIndent\SideShow\vbox to15pt{\endcorner\vfill}\vskip-10pt}}

\def\EndBeginSection{{\vbadness 100000\par\ni%
\global\advance\SideIndent by-1\hskip\SectionIndent\SideShow
\vbox to15pt{\vfill\endbegincorner}}%
\global\advance\SideIndent by1\vskip-10pt}

\def\ShowBeginCorners#1{%
\SideIndentTemp =#1 \advance\SideIndentTemp by-1%
\ifnum \SideIndentTemp>0 %
\vskip-15truept\hbox{\kern 2truept\vbox{\hbox{\begincorner}%
\ShowBeginCorners{\SideIndentTemp}\vskip-3truept}}%
\fi%
}

\def\ShowEndCorners#1{%
\SideIndentTemp =#1 \advance\SideIndentTemp by-1%
\ifnum \SideIndentTemp>0 %
\vskip-15truept\hbox{\kern 2truept\vbox{\hbox{\endcorner}%
\ShowEndCorners{\SideIndentTemp}\vskip 2truept}}%
\fi%
}

\def\BeginSections#1{{\vbadness 100000 \par\ni\hskip\SectionIndent%
\SideShow\vbox to 15pt{\vfill\ShowBeginCorners{#1}}}\global\advance\SideIndent by#1\vskip-10pt}

\def\EndSections#1{{\vbadness 100000 \par\ni\global\advance\SideIndent by-#1%
\hskip\SectionIndent\SideShow\vbox to15pt{\vskip15pt\ShowEndCorners{#1}\vfill}\vskip-10pt}}

\def\EndBeginSections#1#2{{\vbadness 100000\par\ni%
\global\advance\SideIndent by-#1%
\hbox{\hskip\SectionIndent\SideShow\kern-2pt%
\vbox to15pt{\vskip15pt\ShowEndCorners{#1}\vskip4pt\ShowBeginCorners{#2}}}}%
\global\advance\SideIndent by#2\vskip-10pt}




%
%


\def\al{\alpha}
\def\be{\beta}
\def\de{\delta}
\def\ga{\gamma}

\def\la{\lambda}

\def\om{\omega}
\def\si{\sigma}

\def\Ga{\Gamma}





 \def\R{{\hbox{\Bbb R}}}

 \def\R{{\hbox{\Bbb R}}}


\def\ip{\hbox to4pt{\leaders\hrule height0.3pt\hfill}\vbox to8pt{\leaders\vrule width0.3pt\vfill}\kern 2pt}

\def\arr{\rightarrow}

%
%

\def\cases#1{\left\{\eqalign{#1}\right.}
\NormalStyle
\SetNumberingModeSN
\PreventDoubleOn

\long\def\title#1{\centerline{\TitleStyle\ni#1}}
\long\def\moretitle#1{\baselineskip18pt\centerline{\TitleStyle\ni#1}}
\long\def\author#1{\ms\centerline{\AuthorStyle by {\it #1}}}

\long\def\address#1{\ss\centerline{\AddressStyle #1}\par}
\long\def\moreaddress#1{\centerline{\AddressStyle #1}\par}
\def\abstract{\ms\leftskip 3cm\rightskip .5cm\AbstractStyle{\bf \ni Abstract:}\ }
\def\endabstract{\par\leftskip 2cm\rightskip 0cm\NormalStyle\ss}

\SetNumberingModeSN

\def\frac[#1/#2]{\hbox{$#1\over#2$}}
\def\Frac[#1/#2]{{#1\over#2}}
\def\({\left(}
\def\){\right)}
\def\[{\left[}
\def\]{\right]}
\def\^#1{{}^{#1}_{\>\cdot}}
\def\_#1{{}_{#1}^{\>\cdot}}
\def\Label=#1{{\buildrel {\hbox{\fiveSerif \ShowLabel{#1}}}\over =}}
\def\<{\kern -1pt}


\def\ExpandAllCNotes{\long\def\CNote##1{%
\BeginSection
	\Note%
 		##1%
	\endNote%
\EndSection%
}}
\ExpandAllCNotes
%
%
%
%


\def\frame#1{\vbox{\hrule\hbox{\vrule\vbox{\kern2pt\hbox{\kern2pt#1\kern2pt}\kern2pt}\vrule}\hrule\kern-4pt}}

\def\Box to #1#2#3{\frame{\vtop{\hbox to #1{\hfill #2 \hfill}\hbox to #1{\hfill #3 \hfill}}}}


\bib{EPS}{J.Ehlers, F.A.E.Pirani, A.Schild, 
{\it The Geometry of Free Fall and Light Propagation},
in General Relativity, ed. L.OÕRaifeartaigh (Clarendon, Oxford, 1972). 
}

\bib{EPS1}
{M.Di Mauro, L. Fatibene, M.Ferraris, M.Francaviglia, 
{\it Further Extended Theories of Gravitation: Part I },
Int. J. Geom. Methods Mod. Phys. Volume: 7, Issue: 5 (2010), pp. 887-898; gr-qc/0911.2841}

\bib{EPS2}
{L. Fatibene, M.Ferraris, M.Francaviglia, S.Mercadante,
{\it Further Extended Theories of Gravitation: Part II},
Int. J. Geom. Methods Mod. Phys. Volume: 7, Issue: 5 (2010), pp. 899-906; gr-qc/0911.284}

\bib{ELQG}
{L. Fatibene, M. Ferraris, M. Francaviglia,
{\it Extended Loop Quantum Gravity},
CQG 27(18) 185016 (2010); arXiv:1003.1619}

\bib{MGaCou}
{L.Fatibene, M.Francaviglia, S. Mercadante,
{\it Matter Lagrangians Coupled with Connections}
Int. J. Geom. Methods Mod. Phys. Volume: 7, Issue: 5 (2010), 1185-1189; arXiv: 0911.2981}

\bib{catalogo}{ H.Stephani, D.Kramer, M.Mac Callum,
{\it Exact Solutions Of Einstein's Field Equations},
Cambridge University Press (2003) }

\bib{Perlick}{V. Perlick, 
{\it Characterization of standard clocks by means of light rays and freely falling particles}
General Relativity and Gravitation,  {\bf 19}(11) (1987) 1059-1073}

\bib{Faraoni}{T.P. Sotiriou, V. Faraoni,
{\it  $f (R)$  theories of gravity}, (2008); 
arXiv: 0805.1726v2
}

\bib{Capozziello}{S. Capozziello, M. De Laurentis, V. Faraoni
{\it A bird's eye view of $f(R)$-gravity}
(2009); arXiv:0909.4672 
}

\bib{Magnano}{G. Magnano, L.M. Sokolowski, 
{\it On Physical Equivalence between Nonlinear Gravity Theories}
Phys.Rev. D50 (1994) 5039-5059; gr-qc/9312008
}

\bib{S2}{T.P. Sotiriou,
{\it $f(R)$ gravity, torsion and non-metricity},
Class. Quant. Grav. 26 (2009) 152001; gr-qc/0904.2774}

\bib{S3}{T.P. Sotiriou,
{\it Modified Actions for Gravity: Theory and Phenomenology},
Ph.D. Thesis; gr-qc/0710.4438}

\bib{C1}{S. Capozziello, M. Francaviglia,
{\it Extended Theories of Gravity and their Cosmological and Astrophysical Applications},
Journal of General Relativity and Gravitation 40 (2-3), (2008) 357-420.}

\bib{C2}{S. Capozziello, M.F. De Laurentis, M. Francaviglia, S. Mercadante,
{\it From Dark Energy and Dark Matter to Dark Metric},
Foundations of Physics 39 (2009) 1161-1176
gr-qc/0805.3642v4}

\bib{C4}{S. Capozziello, M. De Laurentis, M. Francaviglia, S. Mercadante,
{\it First Order Extended Gravity and the Dark Side of the Universe Ð II: Matching Observational Data},
Proceedings of the Conference ``Univers Invisibile'', Paris June 29 Ð July 3, 2009 
Ð to appear in 2010}

\bib{Olmo}{Olmo Sigh}

\bib{Schouten}{J.A.Schouten,
{\it Ricci-Calculus: An Introduction to Tensor Analysis and its Geometrical Applications},
Springer Verlag (1954)
}

\bib{Bibliopolis}{M.Francaviglia,
{\it Relativistic theories},
Quaderni del CNR-GNFM (Italy, 1988)
}

\bib{Weinberg}{S.Weinberg
{\it Gravitation and Cosmology: Principles and Applications of the General Theory of Relativity},
John Wiley \& Sons (1972)
}

\bib{Landau}{L.D.Landau, E.M.Lifschitz , {\it  The Classical Theory of Fields - Vol 2},
Reed Educational and Professional Publishing (1975)}

\bib{Trautman}{A.Trautman, 
{\it The general theory of relativity}, 
Usp. fiz. Nauk {\bf 89} (1966) 3-37
}


\bib{MCC}{L.Fatibene, M.Francaviglia, S.Mercadante,
{\it Matter Lagrangians Coupled with Connections},
Int. J. Geom. Meth. Mod. Phys. {\bf 7} (2010) 1185-1189}

\bib{Dadhich}{N.Dadhich,
{\it Universal features of gravity and higher dimensions},
in: {\it 13th Regional Conference on Mathematical Physics} Oct. 23-31, 2010 at Antalaya, Turkey
(to appear); arXiv:1105.0988v1 [gr-qc]
}



\def\ubal{\underline{\al}\kern1pt}
\def\obal{\overline{\al}\kern1pt}

\def\ubR{\underline{R}\kern1pt}
\def\obR{\overline{R}\kern1pt}
\def\ubom{\underline{\om}\kern1pt}
\def\obxi{\overline{\xi}\kern1pt}
\def\ubu{\underline{u}\kern1pt}
\def\ube{\underline{e}\kern1pt}
\def\obe{\overline{e}\kern1pt}

\NormalStyle

\title{On a Characterization of Geodesic Trajectories }
\moretitle{and Gravitational Motions}

\author{L.Fatibene$^{a,b}$, M.Francaviglia$^{a,b}$, G.Magnano$^{a}$}

\address{$^a$ Department of Mathematics, University of Torino (Italy)}
\moreaddress{$^b$ INFN - Iniziativa Specifica Na12}

\abstract
We shall here discuss a characterization of geodesics trajectories. 
We shall show that the action of the gravitational field on mass particles can be essentially identified with 
the force that cannot be {\it absolutely} eliminated.
This leads to an alternative formulation of equivalence principle.
\endabstract

\NewSection{Introduction}

It has been noted that gravitational force can be eliminated at a point in spacetime by a class of observers, called {\it free fall observers}.
Some authors thence noted that gravitational field is properly described by tidal forces which in fact cannot be eliminated 
by the choice of the observer; see \ref{Weinberg}, \ref{Landau}, \ref{Dadhich}.

These properties are consequence of  {\it equivalence principle} which claims in fact that mass particles falls along geodesic worldlines. In particular one usually claims that there are free fall observers for which mass particles behave nearly as in special relativity; see \ref{Trautman}, \ref{EPS}.
From a foundational viewpoint there are good reasons to try to characterize worldlines of mass particles
without resorting to special relativity. In fact general relativity (GR) aims to be a fundamental theory and special relativity should be obtained as a week field approximation. 
Hence, in principle, one should not use special relativity essentially in the definition of GR.

We shall hereafter show that mass particle worldlines can be characterized completely in terms of few basic properties.

First,  we know that worldlines are uniquely determinated once one specifies the initial point and the initial {\it velocity} (i.e.~the direction in spacetime). Of course only timelike direction can be used for mass particle worldlines. In any event we expect worldlines to be solutions of ordinary differential equations, of order 2 and Cauchy theorem should hold for it, i.e.~the equation should be in normal form.
In other words we shall start from an equation in the general form
$$
\ddot  q^\mu = f^\mu (s, q, \dot q)
\fl{Eq1}$$
where dots denote the derivatives with respect to the affine parameter $s$ along the curve.
Of course, one should discuss reasonable regularity conditions for the function $ f^\mu (s, q, \dot q)$.

Second, as any equation in GR we expect the equation to be {\it covariant}, i.e.~coordinate independent or, equivalently, diffeomorphism covariant. At least one should describe what happens to the equation when a spacetime diffeomorphism is applied.

Third, since the parametrization of the worldline is unphysical (in the sense that two different parametrizations of the same trajectory represent the same motion of the particle), the equation should be covariant with respect to arbitrary reparametrizations. At least one should describe what happens to the equation when a reprarametrization is applied.

We shall show that there are still many equations obeying these axioms. This is quite an expected result since one would require these properties to hold for any mass particle sujected to an arbitrary force field as well, not when only gravitational field acts.
Hence one needs to characterize pure gravitational motions.

Fourth property will be such characterization. Pure gravitational motions cannot be eliminated {\it absolutely} from the general equation.
Here by {\it absolutely} we mean that one cannot set gravitational field to disappear for any observer and independently of the parametrization used for worldlines at the same time. We shall below show how to mathematically require this property. From a physical viewpoint, for example electromagnetic fields acts on charges but it does not act on neutral particles. As a consequence one could expect there is an equation  for charges in which coupling to electromagnetic field is active and an equation for neutral particles in which there is no interaction with electromagnetic field. When a neutral particle is described the coupling terms to the electromagnetic field is not there in an absolute sense since the particle is neutral for any observer and regardless the parametrization one can use along the worldlines. On the contrary the gravitational interaction is universal. It can be switched off by specific observers but there are always observers for which the gravitational field is there.

We shall show that these four properties in fact characterize completely gravitational interaction and that the equation for mass particles which obeys these properties is necessarily the equation for geodesic trajectories of some connection. We shall also briefly review the issue concerning the observability of such a connection; see \ref{EPS} and \ref{Schouten}. In particular, we shall show that geodesic trajectories are associated to a class of {\it projectively equivalent} connections.

From a fundamental viewpoint one can use this procedure to introduce geodesics in GR to describe mass particle worldlines, without resorting to special relativity.

\NewSection{The Equation for Geodesic Trajectories}

Let us first fix notation. Let us denote by $M$ the spacetime manifold. It is a connected $m$-dimensional smooth manifold on which global Lorentzian metrics exists. Lighrays determine on $M$ a conformal structure $[g]$, i.e.~a class of conformally equivalent Lorentzian metrics. The conformal structure is equivalent to the sheaf of lightcones and it defines {\it lightlike}, {\it timelike} and {\it spacelike} directions in spacetime; see \ref{EPS}, \ref{EPS1}, \ref{EPS2}.

A {\it curve} is a parametrized curve in $M$, i.e.~a map $\ga:I\arr M$ for some interval $I\subset \R$. It is not to be confused with its image
$\ga(I)\subset M$ which is a $1$-dimensional submanifold which we shall called a {\it path}. 

Physically speaking, when two curves are two different parametrizations of the same path then they represent the same motion.
For this reason we define such two curves to be equivalent. 
An equivalence class $[\ga]$ with respect to this relation is called a {\it trajectory} in $M$. A trajectory is characterized by a the path which is shared by all representatives of the trajectory. 

Let us now expand the right hand side of equation \ShowLabel{Eq1} as a formal series of the velocities, i.e.
$$
\ddot  q^\mu = A^\mu (s, q) +  A^\mu_\al (s, q) \dot q^\al +    A^\mu_{\al\be} (s, q) \dot q^\al\dot q^\be +
 A^\mu_{\al\be\ga} (s, q) \dot q^\al\dot q^\be\dot q^\ga + \dots
\fl{Eq2}$$
This can be done whenever the function $f(s, q, \dot q)$ is smooth.

Let now $\Phi:M\arr : q\mapsto q'=Q(q)$ be a spacetime diffeomorphism. 
Then one has
$$
\cases{
&\dot q'^\al= J^\al_\rho \dot q^\rho\cr 
&\ddot q'^\al= J^\al_{\rho\si} \dot q^\rho \dot q^\si + J^\al_\rho \ddot q^\rho\cr
}
\fn$$
where $J^\al_\rho$ and $ J^\al_{\rho\si}$ denotes the Jacobian and Hessian of the diffeomorphism $q'=Q(q)$.

Hence the coefficients  of equation \ShowLabel{Eq2} under the diffeomorphism tranform as
$$
\cases{
&A'^\mu= J^\mu_\nu A^\nu\cr
&A'^\mu_\al=J^\mu_\nu A^\nu_\la  \bar J^\la_\al  \cr
&A'^\mu_{\al\be} =J^\mu_\nu \(A^\nu_{\rho\si} \bar J^\rho_\al  \bar J^\si_\be  -  \bar J^\nu_{\al\be}\)\cr
&A'^\mu_{\al\be\ga}=J^\mu_\nu A^\nu_{\rho\si\la} \bar J^\rho_\al\bar J^\si_\be \bar J^\la_\ga  \cr
&\dots\cr
}
\fn$$
where the bar denotes Jacobians and Hessians of the inverse map $q=\bar Q(q')$.
 Hence to ensure covariance one only needs all coefficients to be tensor fields except the coefficient of order 2.
The transformation rules of the coefficient $A^\nu_{\rho\si}$  
can be easily compared to the transformation rules of a connection $\Ga^\nu_{\rho\si}$ 
to conclude that necessary there exists a connection such that $A^\nu_{\rho\si} = - \Ga^\nu_{\rho\si}$.

Hence the most general covariant equation of the form \ShowLabel{Eq2} is
$$
\ddot  q^\mu  +   \Ga^\mu_{\al\be}  \dot q^\al\dot q^\be= A^\mu   + A^\mu_\al \dot q^\al   +
 A^\mu_{\al\be\ga} \dot q^\al\dot q^\be\dot q^\ga + \dots
\fl{Eq3}$$
understanding that all coefficients $(A^\mu, A^\mu_\al, A^\mu_{\al\be\ga}, \dots)$ are tensorial and $\Ga^\mu_{\al\be}$ is a connection.
Since the equation is insensitive to torsion of $\Ga$ one can restrict without loss of generality to torsionless connections.

Now let us consider an arbitrary reparametrization $q'(s')= q(\phi(s'))$. One has
$$
\cases{
&\dot q'=  \dot q \dot \phi\cr 
&\ddot q'=  \ddot q \dot \phi^2 +  \dot q \ddot \phi\cr
}
\fn$$
and the trasformed equation has new coefficients 
$$
\cases{
&A'^\mu= \dot \phi^2 A^\mu\cr
&A'^\mu_\al = \dot \phi  A^\mu_\al      +  \frac[\ddot \phi/ \dot \phi] \de^\mu_\al\cr
&   \Ga'^\mu_{\al\be}= \Ga^\mu_{\al\be}\cr
&  A'^\mu_{\al\be\ga}=\frac[1/ \dot \phi] A^\mu_{\al\be\ga} \cr
&\dots\cr
}
\fn$$ 
Hence while the connection is invariant with respect to reparametrizations, the other tensorial coefficients are not.
Notice however that $(A^\mu, A^\mu_{\al\be\ga}, \dots)$ transform linearly and if they are set to zero in one parametrization they are zero in any parametrization. Not the same can be said for the coefficient $A^\mu_\al$ which if zero in one parametrization is in general non-zero in the others parametrizations. 
According to our definition of gravitational field, the coefficients $(A^\mu, A^\mu_{\al\be\ga}, \dots)$ can be set to zero {\it absolutely} and they represent non-gravitational interactions. The equation for pure gravitational interactions thence simplifies to
$$
\ddot  q^\mu  +   \Ga^\mu_{\al\be}  \dot q^\al\dot q^\be=  A^\mu_\al \dot q^\al  
\fl{Eq4}$$

Also part of the last coefficient $A^\mu_\al $ can be set to zero absolutely. In fact one can decompose $A^\mu_\al$ as a traceless part $a^\mu_\al$ and a pure trace part, i.e.~$A^\mu_\al= a^\mu_\al + \la \de^\mu_\al$. One can easily show that
$$
\cases{
&\la'  = \dot \phi  \la    + \frac[\ddot \phi/ \dot \phi] \cr
& a'^\mu_\al =  \dot \phi  a^\mu_\al    \cr
}
\fn$$
and the traceless part $a^\mu_\al$ can be also be set to zero absolutely. The equation for gravitational motions further reduces to
$$
\ddot  q^\mu  +   \Ga^\mu_{\al\be}  \dot q^\al\dot q^\be=  \la \dot q^\mu 
\fl{Eq5}$$

\NewSection{Projective Structures on Spacetime}

The equation \ShowLabel{Eq5} coincides with the equation for geodesic trajectories of $\Ga$.
Let us in fact consider a (torsionless) connection $\Ga^\al_{\mu\nu}$ on $M$. A {\it geodesic motion} is a curve locally expressed by
 $\ga: I\arr M: s\mapsto q^\mu(s)$ which satisfies the following equation
 $$
 \ddot q^\al + \Ga^\al_{\mu\nu} \dot q^\mu \dot q^\nu =0
 \fn$$
A trajectory $[\ga]$ is called a {\it geodesic trajectory} for $\Ga$ if one of its representatives is a geodesic motion.
One can easily show that for any representative of a geodesic trajectory $\ga'\in [\ga]$ there exists a function $\la(s)$ such that
locally  $\ga': I\arr M: s\mapsto q'^\mu(s)$  and 
 $$
 \ddot q'^\al + \Ga^\al_{\mu\nu} \dot q'^\mu \dot q'^\nu =\la \dot q'^\al 
 \fn$$
We have hence shown that under very mild regularity conditions and the axioms discussed above the equation \ShowLabel{Eq1} reduces necessarily in the form of this equation.

Let us now define an equivalence relation on the space of all (torsionless) connections. 
Let us say that two connections are {\it projectively equivalent}  when they have the same geodesics trajectories. 
Two connections $\Ga'^\mu_{\al\be}$ and $\Ga^\mu_{\al\be}$ are projectively equivalent if and only if locally there exists a covector $A$
for which
$$
\Ga'^\mu_{\al\be}=\Ga^\mu_{\al\be} + \de^\mu_{(\al}A^{\phantom{\mu}}_{\be)}
\fn$$
If a motion $\ga$ is a representative for a geodesics trajectory for $\Ga$ it is also a geodesic trajectory for $\Ga'$. Hence it obeys the equations
$$
\ddot  q^\mu  +   \Ga^\mu_{\al\be}  \dot q^\al\dot q^\be=  \la \dot q^\mu 
\qquad\qquad
\ddot  q^\mu  +   \Ga'^\mu_{\al\be}  \dot q^\al\dot q^\be=  \la' \dot q^\mu 
\fn$$ 
where we set $\la' = \la + A_\mu \dot q^\mu$.
A class of projectively equivalent connections is called a {\it projective structure} on spacetime; see \ref{Schouten}.
The equation of geodesic trajectories is compatible with the quotient onto projective structures and hence it is attached to a projective structure on spacetime, not to a specific connection. 
In principle one cannot observe a representative of a projective structure just observing worldlines of mass particles.

\NewSection{Conclusions}

We showed that gravitational free fall of mass particles is uniquely characterized by few axioms. 
We know that free fall worldlines are described by a second order normal ordinary differential equation, 
which has to be covariant with respect to spacetime diffeomorphisms and reparametrizations. 
Then gravitational interaction is characterized by the fact that gravitational field cannot be switched off absolutely.
Under these only mild assumptions one can obtain equations of geodesic trajectories for a projective structure on spacetime.

In EPS paper (see \ref{EPS}) one could argue that particle worldlines select a projective structure on spacetime. This is done without resorting to special relativity. In particular there is no reason to assume that the connection describing free fall is Levi-Civita connection of the metric describing lightcones. Extended theories of gravitation, namely $f(R)$ models, in its metric-affine formulation do in fact provide
examples in which free fall turns out to be described by a Weyl connection which is metric but not for the metric originally used to describe lightcones. On the contrary free fall is associated to a conformal metric so that there is a representative of conformal structure which also describe free fall; see \ref{EPS1}, \ref{EPS2}, \ref{MCC}. Although it is well-known that conformal transformations (acting on the metric and leaving the connection unchanged) maps a formalism into an equivalent one, still there is one representative (known as the {\it Einstein frame}) which can be canonically selected; \ref{Magnano}. 

It may be worth noticing that this situation is paradigmatic about how {\it absolute knowledge} arises in relativistic theories. An observer chooses its coordinates and parametrizations to describe worldlines. To write the equation a pair $(\Ga, \la)$ must be chosen. Equivalently any new pair
$(\tilde \Ga^\mu_{\al\be}=\Ga^\mu_{\al\be} + \de^\mu_{(\al}A_{\be)}, \tilde \la = \la + A_\mu \dot q^\mu)$ is equally good for that observer.
This observer dependent setting contains some absolute knowledge about real world together with some convention which is only a characteristic of the observer, namely {\it a gauge}. One absolutely knows that another observers can use another coordinate system, other parametrizations and other pairs $(\Ga', \la')$. However, the new observer is not completely free in its choices if it has to describe the same reality of the previous observer. The connections must be projectively equivalent and the functions $\la$ and $\la'$ are accordingly constrained.
An {\it absolute} description of reality in fact emerges from a set of compatible {\it relative} descriptions. As usual absolute knowledge is encoded in transformation rules as an absolute description of a manifold is encoded in transition functions of an atlas.

\Acknowledgements
We acknowledge the contribution of INFN (Iniziativa Specifica NA12) and the local research project 
{\it Leggi di conservazione in teorie della gravitazione classiche e quantistiche} (2010) of Dipartimento di Matematica of University of Torino (Italy).

We wish to thank M.Ferraris for discussions and comments.

\ShowBiblio

\end